# Mechanisms of Afterglow and Thermally Stimulated Luminescence in UV-irradiated InP/ZnS Quantum Dots

S.S. Savchenko[1,]*, A.O. Shilov[1], A.S. Vokhmintsev[1], I.A. Weinstein[1, 2]

[1]*NANOTECH Centre, Ural Federal University, Yekaterinburg, 620002 Russia*
[2]*Vatolin Institute of Metallurgy, Ural Branch, Russian Academy of Sciences, Yekaterinburg, 620108 Russia*
*E-mail: s.s.savchenko@urfu.ru

Indium phosphide-based quantum dots (QDs) are a potential material for designing optoelectronic devices, owing their adjustable spectral parameters over the entire visible range, as well as their high biocompatibility and environmental safety. Concurrently, they exhibit structural defects, the rectification of which is crucial for enhancing their optical properties. The present work explores, for the first time, the low-temperature afterglow (AG) and spectrally resolved thermally stimulated luminescence (TSL) of UV-irradiated colloidal core/shell InP/ZnS QDs in the range of 7–340 K. It is shown that, when localized during irradiation and released after additional stimulation, charge carriers recombine involving defect centers based on indium and phosphorus dangling bonds. The mechanisms of the observed luminescent phenomena can be caused by both thermal activation and tunneling processes. By means of the initial rise method, the formalism of general-order kinetics, and the analytical description using the Lambert $W$ function, we have analyzed the kinetic features of possible thermally stimulated mechanisms. We have also estimated the energy characteristics of appropriate trapping centers. A low rate of charge carriers recapture is revealed for InP/ZnS QDs. Active traps in nanocrystals of different sizes are characterized by close values of activation energy in the 26–31 meV range. The current paper discloses new horizons for exploiting TSL approaches to study the properties of local defective states in the energy structure of colloidal QDs, which can contribute to the development of targeted synthesis of nanocrystals with tunable temperature sensitivity for optoelectronic and sensor applications.

Keywords: afterglow, spectrally resolved thermoluminescence, Lambert $W$ function, colloidal quantum dots, core/shell, indium phosphide, interface, dangling bonds, traps, tunneling, activation energy, kinetic order, retrapping ratio.

# Introduction

The unique optical and electronic properties of semiconductor colloidal quantum dots (QDs) have motivated researchers to engage in intensive investigations over more than four decades [1–4]. Such zero-dimensional objects spark a great interest in how their size affects the QDs' energy structure, absorption and emission wavelengths, as well as in the possibility of creating biologically compatible and chemically stable multilayer QDs with high quantum efficiency, which necessitates their application in the field of labeling, optoelectronics, and nanophotonics [5–9].

The large surface-to-volume ratio inherent in nanoscale QDs results in a developed surface, essentially changing their optical properties [10]. It is known that the radiative relaxation of excited excitons governs the photoluminescence (PL) spectrum of QDs. However, chalcogenide- and phosphide-based QDs demonstrate additional long-wavelength luminescence bands caused by structural defects. Compared to the exciton luminescence bands, these are characterized by a broader half-width, with the appropriate recombination processes proceeding with slower decay kinetics [11–16]. In addition, defect states can serve as charge carrier traps [17–22], manifesting themselves in the processes of temperature quenching of the PL of CdSe- [23, 24], CdS- [25], CdS/ZnS- [26], $Ag_2S$- [27, 28], $Ag_2S/SiO_2$- [29, 30], PbS-, $PbS/SiO_2$- [31], and InP/ZnS-QDs [14, 16].

For example, the authors of the paper [24] detected an additional integral luminescent response of the exciton 2.12 eV and defect 1.52 eV emission of CdSe nanoplates when measuring the PL quenching curve. The samples were subjected to heating and continuous laser excitation within the 3.06 eV band. This approach made it possible to identify two types of traps that are obliquely involved in radiative recombination processes and to reconstruct their Gaussian distributions for the density of defect states with energies of 100 and 280 meV and half-widths of 50 and 60 meV, respectively. Simultaneous photo- and thermal stimulation was also used for testing PbS QDs stabilized by various passivators and $PbS/SiO_2$ core/shell structures [31]. Besides, the temperature curves of the luminescence intensity were measured in cooling-heating cycles in the range of 80–350 K at a rate of 0.05 K/s, uncovering two types of traps with a depth of 0.17 and 0.25 eV. It is shown that shell growth enables one to get rid of deep traps, which evidences their interface nature. Thus, simultaneous optical and thermal stimulation dominates in the currently available publications, leaving aside the recorded signal's spectrum resolution. Such a circumstance only indirectly supports the participation of charge carrier traps in the luminescence mechanisms of colloidal QDs.

Existing methods of surface passivation, including the creation of heterostructures with one or more shells, fall short of completely addressing the issue of traps [32–34]. This means that a comprehensive study of the energy defect levels in QDs, as well as the mechanisms of capture and release of charge carriers is still relevant. Furthermore, standard methods of thermal activation spectroscopy are believed to be unsuitable for examining QDs due to the low density of allowed traps' energy states and, consequently, the low intensity of thermally stimulated luminescence (TSL) [23]. However, there are research projects on CdS- [35, 36], CdSe- [37], and $CdS_{1-x}Se_x$- [12] semiconductor nanocrystals in glass matrices irradiated with ultraviolet or X-ray radiation using the TSL method.

So, the work [36] reports on CdS-doped sodium-lime glass irradiated at 77 K by photons with energies of 6 and 3.5 eV to match the regions of intrinsic absorption of glass and nanocrystals. In this case, the TSL curves were recorded during heating at a rate of 0.1 K/s in the spectral range of intrinsic luminescence of glass and impurity luminescence of QDs. The coincidence of the curves measured can be explained by the capture of electrons by deep centers in the glass with their subsequent radiative recombination during heating. Since the QDs' traps acted only as luminescence centers, information on their spectral-kinetic characteristics was then impossible to extract.

The TSL curves were also built for CdSe QDs in aluminoborosilicate glass in the temperature range of 300–775 K [37]. Irradiation was performed with X-rays, and the integral luminescent signal was recorded at a heating rate of 0.1 K/s. As a result of thermal activation of traps in the glass matrix, the samples produced a peak, whatever the irradiation times, observed in the range of 50–300 °C and attributed to the QD luminescence.

Thermally stimulated luminescence was recorded for $CdS_{1-x}Se_x$ nanocrystals in borosilicate glass irradiated with X-rays [12]. When heated at a rate of 0.4 K/s, the samples yielded a TSL signal within the range of 650–850 nm. Comparative analysis of the PL and TSL spectra showed that irradiation with X-rays can initiate a change in the energies of radiative transitions, as a consequence of ionization of nanocrystals. All the TSL curves for samples of different composition and size maintained a maximum at 360–370 K. Summarizing the outcomes of the aforementioned scientific inquiries that resort to the direct TSL method [12, 36, 37], it can be argued that the activated traps are tied with the solid matrix rather than with the energy states in QDs. Similar investigations of the luminescent response for colloidal nanocrystals under thermal activation have not been carried out.

Among the materials used for the fabrication of QDs, cadmium (Cd)- and lead (Pb)-based nanocrystals are the most thoroughly examined regarding the passivation of surface defects [34, 38]. However, their usage is limited due to the possible harmful impacts on human health and the environment. Accordingly, alternative materials that contain no heavy metal ions are needed to create nanophotonic and optoelectronic devices [39]. The most promising option is InP-based QDs due to the ability to adjust their spectral parameters over the entire visible range at the expense of the quantum size effect and their inherent high biocompatibility and environmental safety [16, 40]. It is known that controlling the defects based on the dangling bonds of indium and phosphorus in InP QDs remains a highly complicated task [41, 42]. Moreover, the available literature lacks data on the study of the properties of defect states and mechanisms of capture and release of charge carriers using TSL and afterglow (AG) analysis in such QDs. Recently, we reported briefly on the observation of thermally stimulated luminescence in InP/ZnS upon heating [43]. The current paper unveils, for the first time, the comprehensive analysis of the afterglow and spectrally resolved TSL in the temperature range of 7–340 K of UV-irradiated InP/ZnS core/shell QDs. It also determines the spectral and kinetic characteristics as well as possible mechanisms of the thermally activated processes.

## Samples and Methods

In this work, we have explored two ensembles of water-soluble InP/ZnS QDs synthesized by Applied Acoustics Research Institute (NIIPA, Dubna, Russia) using safe precursors of the aminophosphine type [44]. Previously, we have investigated in-depth the structural characteristics and temperature impacts in the optical absorption and photoluminescence processes for these QD samples [14, 16, 45–47]. InP nanocrystals are coated with a ZnS shell and stabilized with modified polyacrylic acid. The parameters of the first exciton absorption band, 2.1 nm and 11.1% for the QD-1 and 2.3 nm and 17.4% for the QD-2, respectively, underlaid the estimation of the maximum and relative half-width of the size distribution [47]. The PL quantum yield at room temperature was determined relative to the reference sample Rhodamine 6G [48]. The values obtained were 27% for the QD-1 and 10% for the QD-2.

To measure the AG and TSL signal, we used QD samples as films deposited on 1x1 cm and 1 mm thick quartz substrates by evaporating droplets of colloidal solutions in a volume of 120 μl. The luminescent signal was recorded using a Shamrock SR-303i-B spectrograph and a Newton$^{EM}$ DU970P-BV-602 CCD matrix manufactured by Andor in the electron multiplication mode. Such equipment made it possible to amplify the signal by about 100 times. The measurements included cooling the matrix to a temperature of -80 °C. The data were collected in the full vertical binning (FVB) mode. A Janis CCS-100/204N helium cryostat equipped with a LakeShore Model 335 controller provided the temperature control for the samples.

The samples were preliminarily irradiated with a deuterium lamp LD (D) through a UFS-1 filter at 7 K for 10 minutes. After switching off the excitation, it took 40 minutes to record the luminescent signal intensity. The detector exposure time in this mode amounted to 0.995 s. Further, to gain the TSL response, the samples were linearly heated from 7 to 340 K at a rate of 10 K min$^{-1}$. An experiment was also conducted in which heating of the sample was started immediately after irradiation. The integration time of the luminescent signal by the CCD matrix when recording one spectrum was 3 s, which corresponded to a change in the nanocrystal temperature by 0.5 K. The experimental data obtained were plotted as three-dimensional (3D) dependencies along the coordinate axes of temperature/time, wavelength, and luminescence intensity. The paper discusses the temperature/time cross-sections of the these dependencies known as TSL/AG spectra, as well as the spectral sections that correlate to the TSL temperature/AG decay curves. When analyzing spectral dependencies in the energy scale, a corresponding correction was performed [49]. All of the above measurements were also taken for the original quartz substrates. In this case, with the irradiation and stimulation parameters specified, no luminescent signal was detected.

## Results and discussion

Figure 1 illustrates the experimental 3D-dependencies of the $I_{AG}$ luminescence intensity on time and wavelength for the colloidal InP/ZnS QDs studied. As can be seen, the ensembles with different average nanocrystal sizes exhibit a prolonged afterglow upon the cessation of UV irradiation at a constant temperature of 7 K. The spectral ranges corresponding to the $I_{AG}$ peaks are 560–660 nm for the QD-1 and 600–700 nm for the QD-2.

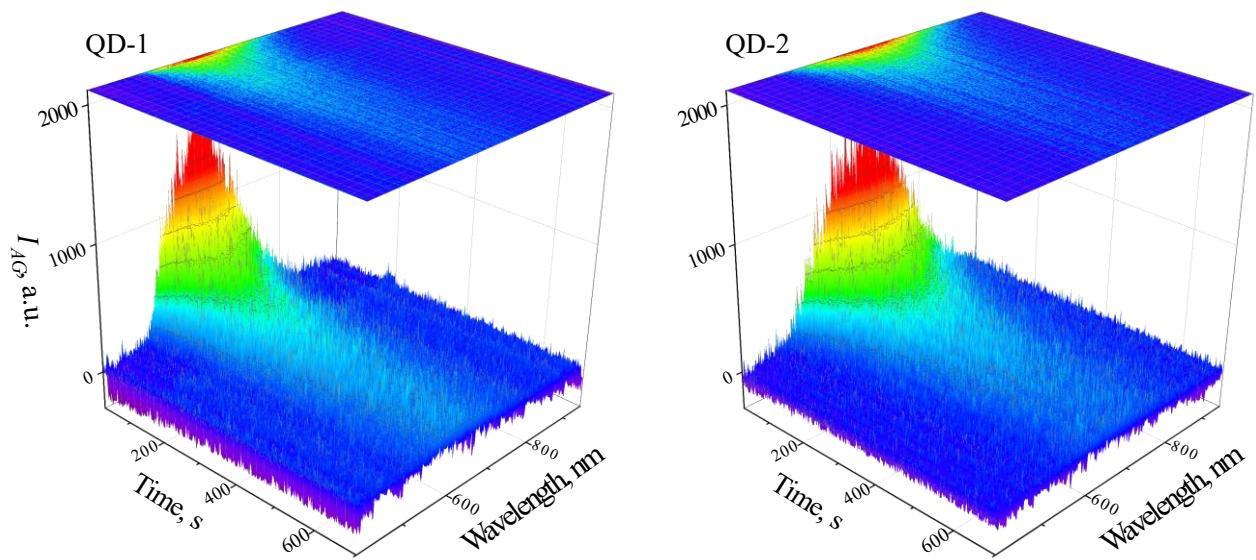

**Figure 1.** Dependence of the afterglow intensity of colloidal InP/ZnS QDs on time and wavelength at $T = 7$ K.

The InP/ZnS QDs afterglow spectra for different delay durations $t_d$ after switching off the irradiation were obtained by averaging the intensity over a period of $t_d \pm 5$ s. Figures 2a and 2c exemplify these outcomes. The intensity maximum for the QD-1 is recorded at 1.97 eV; its half-width amounts to 0.55 eV. Whereas for the QD-2, these values attain 1.78 eV and 0.52 eV, respectively. Whatever $t_d$, the shape of the spectra remains unchanged, as clear from the appropriate insets. The observed emission decays protractedly, hinting at the existence of active traps involved in luminescence processes. Figures 2b and 2d outline the normalized time-dependencies of the afterglow intensity in different spectral ranges indicated by the shaded areas in Figures 2a and 2c. The solid lines provide the experimental data approximation within different models [50, 51]:

$$I_{AG}(t) = A_1 \exp\left(-\frac{t}{\tau_1}\right) + A_2 \exp\left(-\frac{t}{\tau_2}\right), \quad (1)$$

$$I_{AG}(t) = A_3 (t + t_0)^{-p}, \quad (2)$$

where $I_{AG}$ is the afterglow intensity at time $t$, expressed in relative units; $A_i$ are coefficients, also in relative units; $\tau_i$ denotes the decay time of the appropriate component, s; and $p$ is the exponent. The obtained parameter values are listed in Table 1.

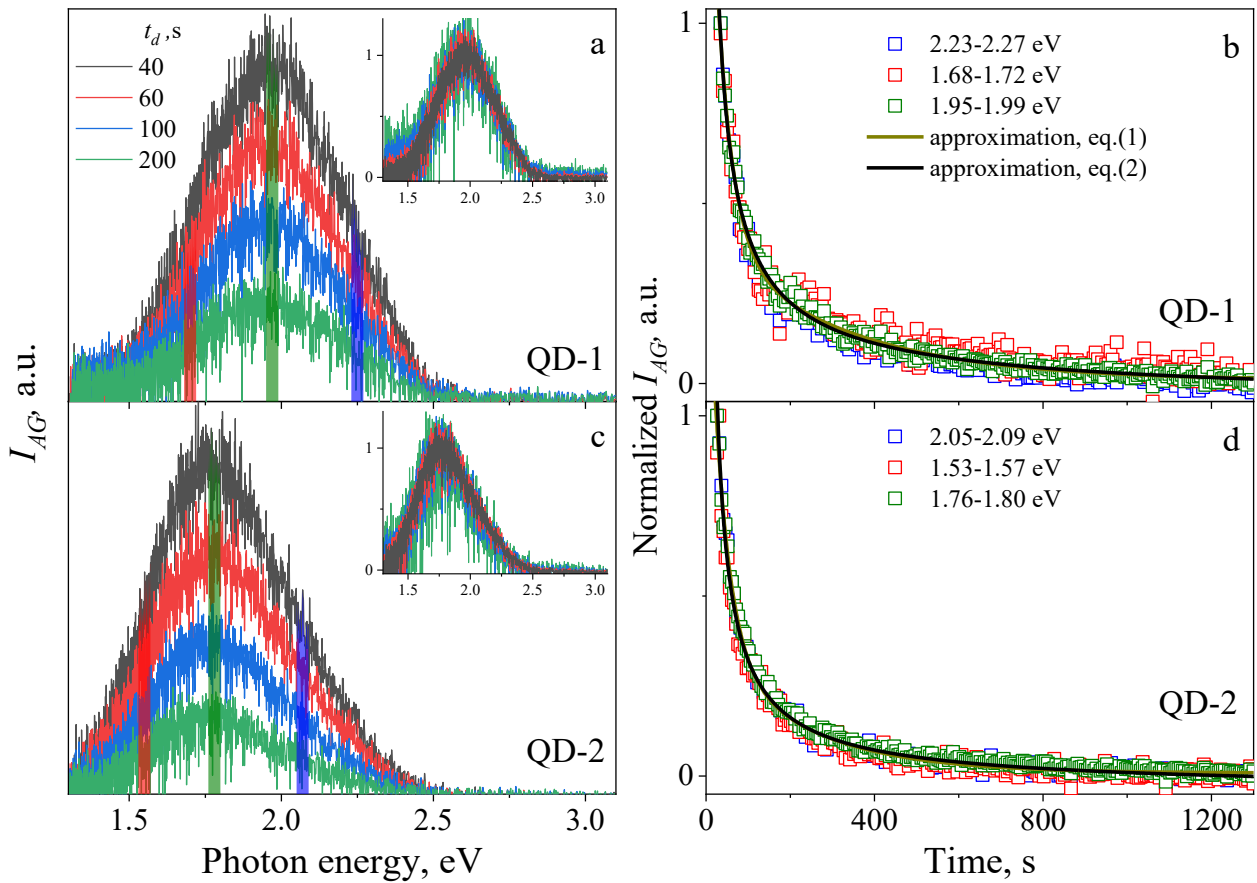

**Figure 2.** Afterglow spectra of InP/ZnS QDs at 7 K for different delay durations $t_d$ (a,c) and afterglow decay curves for the spectral ranges indicated by shaded areas (b,d). The insets display the normalized spectra.

When describing the decay curves by the two-exponential model according to the expression (1), the values of $\tau_1$ and $\tau_2$ match different spectral ranges for the QD-1 and QD-2 samples. The fast component has a feature as a time constant of 37 s for both samples, while the slow one is 315 s for the QD-1 and 225 s for the QD-2. Obviously, the processes mentioned above do not pertain to direct radiative recombination of nanosecond-living excitons [11, 52]. As will be shown below, defect states are involved in the luminescence observed. To describe the long-lasting afterglow decay kinetics, several components need to be utilized. This circumstance may indicate various types of traps/recombination centers or their energy distribution in the QDs examined [53]. It should be emphasized that the QDs' traps activate at a temperature of 7 K, equivalent to a thermal energy of less than 1 meV.

**Table 1.** Spectral and kinetic parameters of afterglow and thermally stimulated luminescence of colloidal InP/ZnS QDs

| Sample | $E_x$, [14] | $E_d$, [14] | $E_{AG-TSL}$, | $H_{AG-TSL}$, | $\tau_1$, | $\tau_2$, | $p$ |
|---|---|---|---|---|---|---|---|
| | ± 0.02 eV | | ± 0.05 eV | | ± 1 s | ± 5 s | ± 0.01 |
| QD-1 | 2.36 | 2.04 | 1.97 | 0.55 | 37 | 315 | 0.73 |
| QD-2 | 2.16 | 1.80 | 1.78 | 0.52 | 37 | 225 | 0.82 |

The power law (2) explains well the afterglow decay kinetics. The exponent of about 1 evidences the release of traps by athermal tunneling of captured carriers [51, 54–58]. It is known that this process takes place exclusively between isoenergetic states [51]. The above statement suggests that localized levels of traps and luminescence centers in InP/ZnS nanocrystals are caused by the same structural defects. Considering the QD size of several nanometers, the contribution of the tunneling mechanism to the trap emptying processes proves to be justified. Furthermore, an analysis of the temperature quenching of the exciton and defect-related PL in InP/ZnS accounting for the activation energy distribution [14] also corroborates the tunneling phenomenon.

The samples were held at 7 K for 40 min before being linearly heated to 340 K at a rate of 10 K·min$^{-1}$. Figure 3 depicts the experimental 3D-dependencies of the thermally stimulated luminescence intensity $I_{TSL}$ on temperature and wavelength for the colloidal InP/ZnS QDs. The TSL spectrum manifests itself as a broad band with a single peak addressing the range of 135–145 K on the temperature scale. On the spectral scale, the $I_{TSL}$ maxima are observed in the same ranges as for the afterglow.

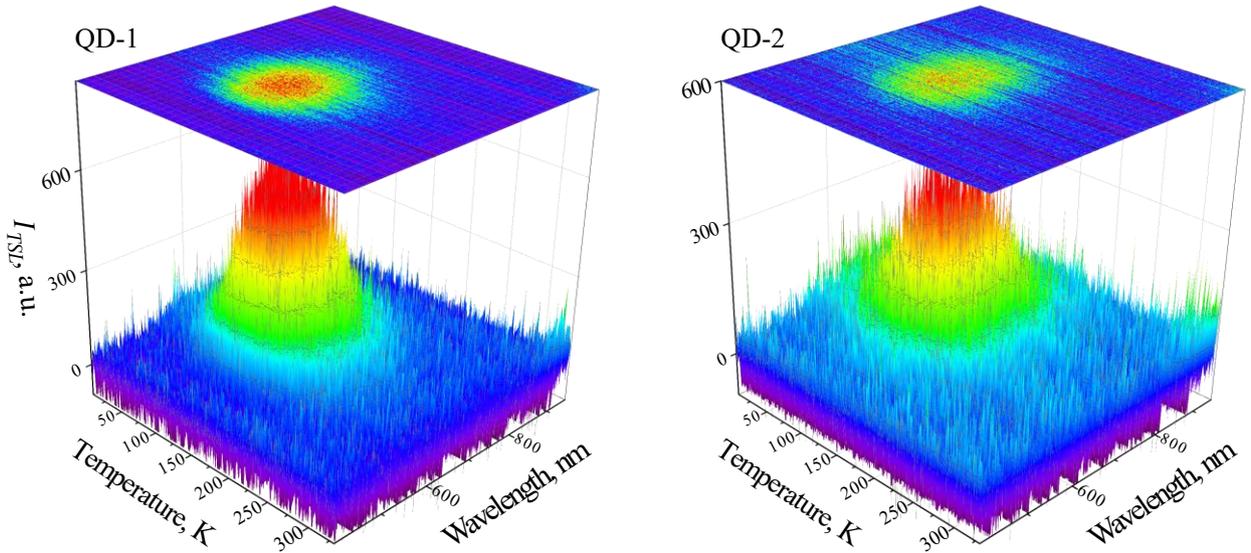

**Figure 3.** Spectrally resolved TSL of colloidal InP/ZnS QDs.

The TSL luminescence spectra at different temperatures were obtained by averaging the intensity in the range of $T \pm 5$ K (see Figures 4a and 4c). The signal intensity increases monotonically with rising $T$, reaches a maximum and gradually drops to zero with additional heating. In this case, the shape of the spectra remains unchanged, as plainly illustrated in the relevant insets. Thus, the samples exhibit the luminescence at different heating temperatures owing to the participation of the same centers. Figures 4b and 4d compare the TSL spectra with both the afterglow and the PL signal measured previously at 6.5 K [14]. As can be seen, the TSL and AG spectra are mutually compatible for each of the samples, sharing the values of their intensity maximum position $E_{AG-TSL}$ and the half-width $H_{AG-TSL}$ (refer to Table 1). This assessment can serve as additional confirmation of the fact that a tunneling mechanism makes it felt in the recombination responsible for the QDs' persistent afterglow [55, 59, 60]. In the process, the position of these bands is in good agreement with the maxima $E_d$ of the defect-related luminescence of the InP/ZnS QDs (see Table 1), indicating a common nature of the recombination centers involved in the luminescent processes. As for the QD-2 having a larger size, their AG and TSL spectra, as well as the defect-related component in the PL spectrum, are shifted to the lower-energy region. It should be noted that these spectra lack the $E_x$ band that is characteristic of exciton luminescence under optical excitation (Figure 4). At room temperature, the defect-related luminescence of the QD-1 and QD-2 is quenched and not detected in the PL spectra of InP/ZnS [14].

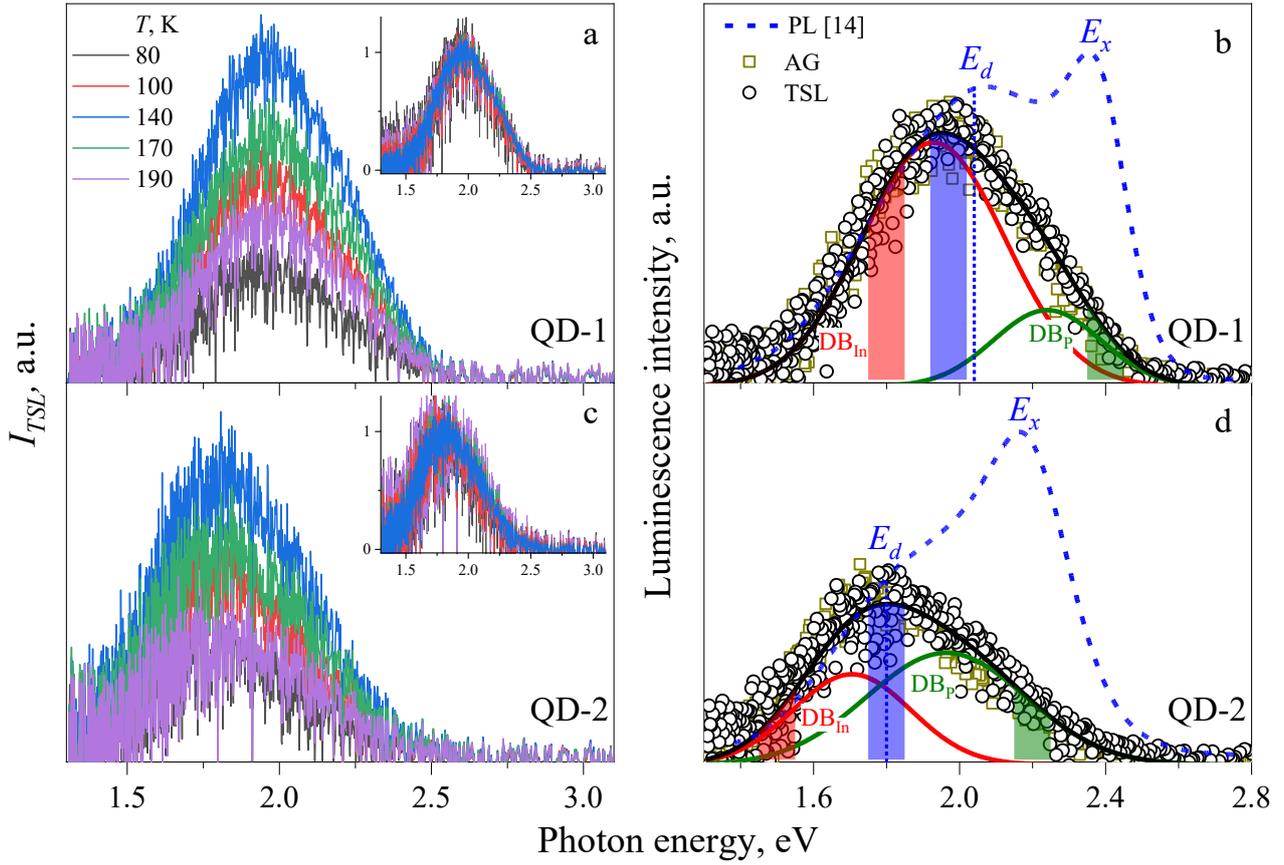

**Figure 4.** InP/ZnS TSL spectra at different heating temperatures of the samples (a,c). Analysis of experimental data on afterglow (square symbols), photo- (dashed line) and thermally stimulated (round symbols) luminescence of QDs (b,d). The black line is the resulting approximation of the AG/TSL spectra when decomposed into two Gaussian components (red and green lines).

Previously, the works [41, 61] theoretically discussed how various defect states affect the electronic and optical properties of InP and InP/ZnS QDs within the density functional method. It was shown that due to the dangling bonds of indium $DB_{In}$ and phosphorus $DB_P$ atoms on the core surface, gap donor and acceptor levels are formed in the energy structure of the QDs, respectively. The depth of the defect levels is higher for $DB_{In}$ than that for $DB_P$ and does not depend on the size of the nanocrystals. Furthermore, substitution defects $Zn_{In}$ (zinc atoms in the crystalline positions of indium) can arise in the core.

The shape of the afterglow and TSL spectra can be described with high accuracy by two Gaussian components depicted in Figures 4b and 4d. For the QD-1, the maximum position and the half-width of the first component are 1.93 eV and 0.46 eV, and for the second component – 2.24 eV and 0.35 eV, respectively. For the QD-2, the aforementioned parameters for the low-energy component reach the values of 1.71 eV and 0.38 eV, and for the high-energy component – 1.96 eV and 0.51 eV. It is known that, as QDs' size decreases, the contribution of the defect-related band to the integrated photoluminescence spectrum of InP/ZnS goes up [11, 14, 16, 62]. This happens due to an enlargement of the proportion of atoms in the interface region relative to the atoms inside the core. During the current experiments, this effect manifests itself in a higher intensity of the TSL response (by a factor of 2) for the QD-1 sample. Based on the modeling findings [41, 61] and interpretation of the experimental data, the high- and low-energy components of the AG and TSL spectra can be assumed to be caused by transitions involving the $DB_P$ and $DB_{In}$ levels, respectively (see Figures 4b and 4d).

Having analyzed the spectral composition of the TSL glow, we dig into the kinetics of the captured charge-carrier release processes. The spectral resolution of the signal allows conducting an analysis over various ranges that correspond to radiative recombination predominantly through structural defects of one type, as shown in Figures 4b and 4d by shaded areas. Thus, the low-energy ranges (marked in red) are mainly linked with the $DB_{In}$-based luminescence centers, whereas the high-energy ones (marked in green) are due to recombination involving the $DB_P$ levels. The intervals marked in blue match the region of maxima in the TSL spectra. The temperature curves plotted for the QD-1 and QD-2 by averaging $I_{TSL}$ for the specified energy ranges are presented in Figures 5a and 5c. Obviously, at $T < 75$ K, all of the curves contain a constant signal due to the prolonged afterglow analyzed above. TSL grows monotonically as the temperature rises, reaching its peak intensity between 135 K and 145 K before diminishing to background values at room temperature. The temperature curves in the insets of Figures 5a and 5c are normalized. It is worth stressing that the shape of the curves for the spectral ranges at hand coincides, and a single TSL peak appears at a temperature of about 140 K. Thus, we can suggest that the emptying of the same traps triggers the TSL glow of the InP/ZnS QDs in all spectral ranges.

The kinetic parameters of the delocalization processes of captured carriers were unraveled within the phenomenological one-trap and one-recombination center model (OTOR) [63]. In Figures 5b and 5d, the symbols designate the TSL temperature curves in the region of spectral maxima, and the solid lines indicate the experimental data approximations within different approaches. For comparative analysis the parameter values were estimated using the initial rise method [63]:

$$I_{TSL}(T) \propto \exp\left(-\frac{E_A}{kT}\right), \tag{3}$$

the general-order kinetics equation with discrete energy levels [63–66]:

$$I_{TSL}(T) = s''n_0 \exp\left(-\frac{E_A}{kT}\right)\left(1 + \frac{s''(b-1)}{\beta}\int_{T_0}^{T}\exp\left(-\frac{E_A}{k\theta}\right)d\theta\right)^{\frac{b}{1-b}}, \tag{4}$$

a semi-analytical solution based on the Lambert $W$ function for the OTOR model [67–69]:

$$I_{TSL}(T) = \frac{NR}{(1-R)^2} \frac{s\exp\left(-\frac{E_A}{kT}\right)}{W[e^z] + W[e^z]^2} \text{ for } R = A_n/A_m < 1,$$

$$z = \frac{1}{c} - \ln(c) + \frac{s}{(1-R)\beta} \int_{T_0}^{T} \exp\left(-\frac{E_A}{k\theta}\right) d\theta,$$

$$I_{TSL}(T) = \frac{NR}{(1-R)^2} \frac{s\exp\left(-\frac{E_A}{kT}\right)}{W[-1,-e^{-z}] + W[-1,-e^{-z}]^2} \text{ for } R = A_n/A_m > 1, \qquad (5)$$

$$z = \frac{1}{|c|} - \ln(|c|) + \frac{s}{(1-R)\beta} \int_{T_0}^{T} \exp\left(-\frac{E_A}{k\theta}\right) d\theta,$$

$$c = \frac{n_0}{N}\frac{1-R}{R}.$$

where $I_{TSL}$ is the TSL intensity, a.u.; $T_0$ is the initial temperature, K; $\beta$ is a sample heating rate, K/s; $k$ is the Boltzmann constant, eV/K; $E_A$ is the activation energy, eV; $b$ is the kinetic order; $s$ is a frequency factor, s$^{-1}$; $s''$ is the effective frequency factor, s$^{-1}$; $n_0$ is the initial concentration of charge carriers captured by traps, cm$^{-3}$; $N$ is the concentration of available traps, cm$^{-3}$; $A_n$ is an electron retrapping coefficient, cm$^3$/s; $A_m$ is an electron recombination coefficient, cm$^3$/s; and $R=A_n/A_m$ is the dimensionless quantity of retrapping ratio.

Figures 5b and 5d illustrate the experimental data approximation for the QDs, utilizing Eqs. (2) and (3), which are highlighted by red and green lines, respectively. The general-order kinetics model enables one to describe them with a maximum accuracy as $b \to 1$, indicating the dominance of the first-order kinetics processes. The description within a semi-analytical solution based on the Lambert $W$ function includes the retrapping ratio $R \ll 1$. Therefore, both approaches point to a low rate of the carrier recapture in the InP/ZnS QDs. The insets reflect the data for estimating the activation energy of traps by the initial rise method in Arrhenius coordinates. The obtained parameter values are summarized in Table 2

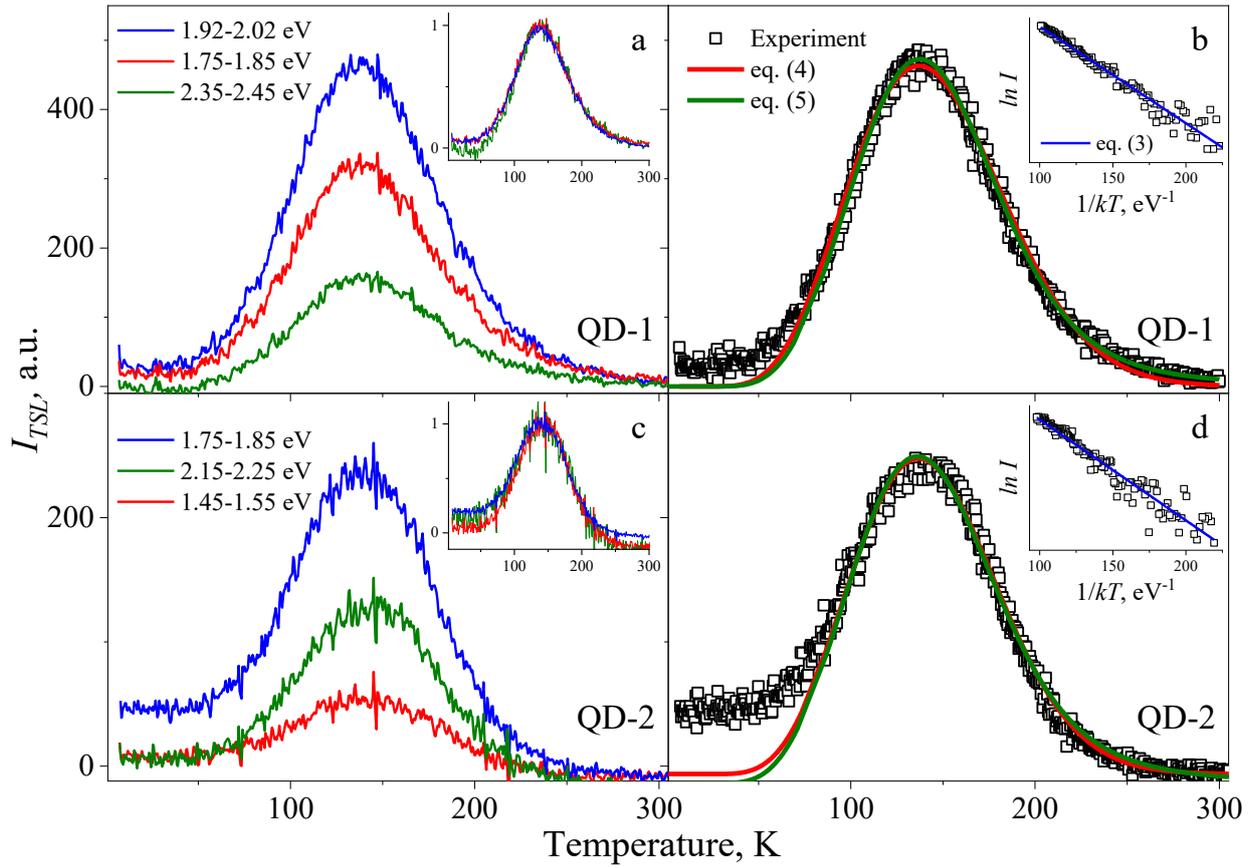

**Figure 5**. TSL glow curves of colloidal InP/ZnS QDs in different spectral ranges (a,c), indicated in Figures 4b and 4d by shaded areas of the appropriate color. The insets show the normalized data. Analysis of the TSL curves within different approaches (b,d). Symbols are experimental data, the red and green lines denote approximations according to models (4) and (5). The inset shows the data in Arrhenius coordinates, the blue line is an approximation in accordance with Eq.(3).

**Table 2.** Parameters of trap release processes in UV-irradiated InP/ZnS colloidal quantum dots

| Sample | $b$, ± 0.02 | $s''$, ± 0.01 s$^{-1}$ | $s$, ± 0.01 s$^{-1}$ | $E_A$, ± 2 meV | | | $R$ |
|---|---|---|---|---|---|---|---|
| | Eq. (4) | (4) | (5) | (3) | (4) | (5) | (5) |
| QD-1 | 1.05 | 0.04 | 0.04 | 26 | 29 | 31 | <0.07 |
| QD-2 | 1.00 | 0.04 | 0.04 | 28 | 26 | 30 | <0.003 |

It is evident that the values of $E_A$ are close in magnitude for different samples and, taking into account an error, almost coincide when using different approaches to describe the experimental TSL data. On the one hand, this outcome uncovers a common nature of the defect trapping centers in the QDs tested. On the other hand, this may also stand for the fact that the model of independent discrete trapping levels has certain limitations and that the approach exploited is not sensitive enough to quantum size effect. It is known that a spread in size, shape, stoichiometry, etc. are inherent to InP/ZnS QDs in an ensemble. As a consequence, this leads to a high level of static structural disorder, which manifests itself in inhomogeneous broadening of optical bands and in the PL temperature quenching mechanisms [14, 16, 46, 47]. Also, the spread is expected to affect the parameters of the relevant traps [63].

A comparative analysis of the experimental findings obtained for the irradiated InP/ZnS linearly heated after holding at 7 K for 0 min and 40 min validates the presence of a distribution of trap parameters. Figure 6a shows the 3D-dependence of the luminescence intensity on temperature and wavelength for the QD-1, recorded during heating immediately after irradiation. It is evident that, unlike the case of holding at 7 K for 40 minutes, the obtained curve responded by two temperature maxima. The spectral characteristics of the luminescence correspond to those given in Table 1. The symbols in Figure 6b trace the experimental time (temperature)-dependence of the luminescence intensity in the range of 1.92–2.02 eV; the highest signal is observed at low $T$. The samples heated to 50 K lose their luminescence intensity; however, it enhances to a maximum at 120 K. Afterwards, the signal value drops to background values at room temperature. As can be seen in Figure 6b, the solid line aligns with the approximation derived from Eq.(1) for a monotonically descending segment in the low-temperature region. The predicted parameters are consistent with the afterglow decay kinetics data for the QDs previously examined at 7 K; refer to Table 1.

The research results testify that the nature of the luminescence intensity decline is almost temperature-independent. One could suggest that tunnel recombination plays a substantial role in the mechanisms underlying the low-temperature luminescent processes [55]. To extract the thermally stimulated contribution observed with rising $T$ against the decaying afterglow background, we have excluded the approximating curve (see Figure 6b) from the experimental one. The TSL curve thus plotted is designated in Figure 6c by a black line. Distinctly, $I_{TSL}$ begins to grow at lower temperatures as compared to that during heating after the sample holding at 7 K for 40 minutes (red line). The $I_{TSL}$ maximum occurs at 130 K versus 140 K, with the high-temperature part of the TSL signal remaining virtually unchanged. This stands for the distribution in energy of active capture centers. With the 40-minute holding period at 7 K, thermally stimulated and tunneling processes give rise to the emptying of some traps with a lower activation energy. This circumstance explains the long-lasting afterglow detected in the irradiated QD samples, bringing about a shift of both the $I_{TSL}$ maximum and the TSL low-temperature wing to the higher temperature region upon further heating. Thus, in the future, the distribution of energy and kinetic parameters for active capture centers and charge carrier recombination can be treated as serious academic inquiries into thermally stimulated luminescence mechanisms in ensembles of colloidal quantum dots.

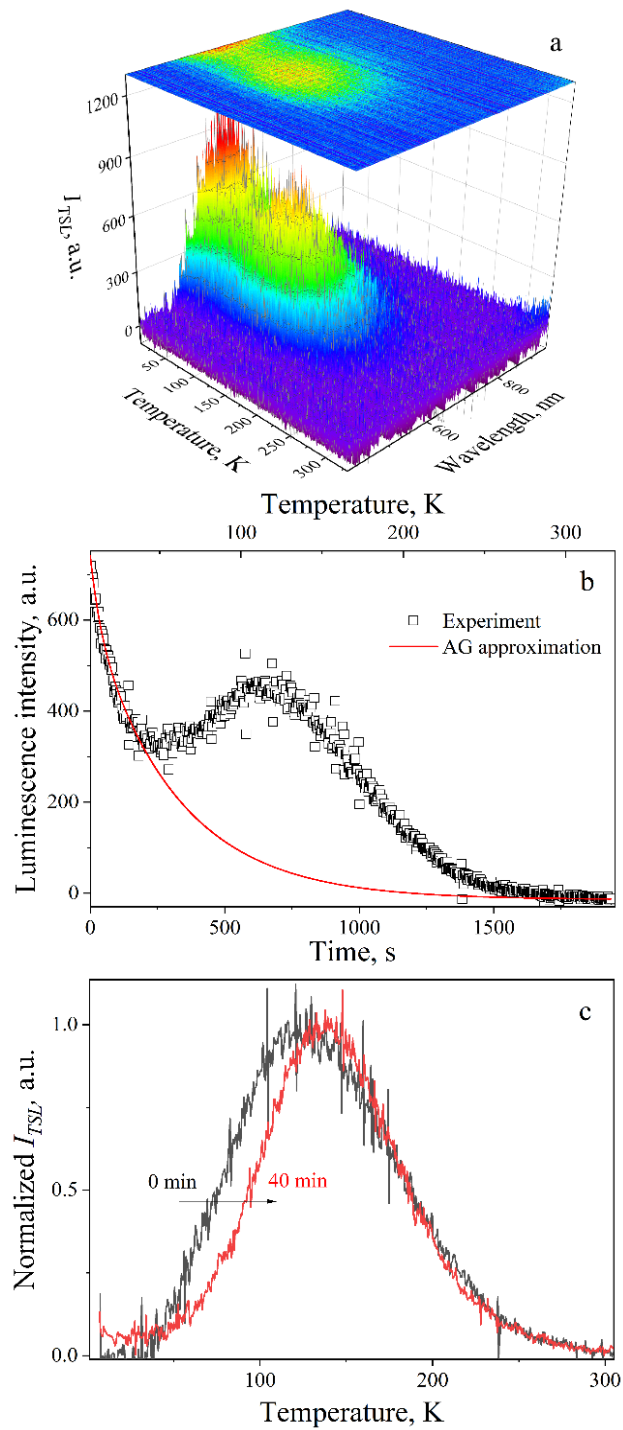

**Figure 6.** Spectrally resolved TSL of QD-1 upon linear heating immediately after irradiation (a). Luminescence intensity versus time (temperature) in the spectral maximum region (b), solid line – approximation according to Eq.(1). TSL temperature curves for 0 minutes (black line) and 40 minutes (red line) of holding at 7 K after switching off UV irradiation (c).

## Conclusion

The experimental peculiarities and regularities of spectrally resolved thermally stimulated luminescence in colloidal InP/ZnS QDs with average sizes of 2.1nm (QD-1) and 2.3 nm (QD-2) in the range of 7–340 K after UV irradiation were studied for the first time, as well as the prolonged afterglow processes in the samples at 7–50 K. The afterglow and TSL spectra are found to be consistent with each other and conform to a spectral band for defect-related photoluminescence in InP/ZnS. These recombination processes appear to be caused by defects based on dangling bonds of indium and phosphorus atoms in the core/shell interface region. Both the two-component exponential model (1) and the power law (2) offer an explanation of the afterglow decay kinetics at 7 K, which may reveal thermally activated and tunneling trap emptying mechanisms. The fast component is characterized by a time constant of 37 s for both samples, whereas the slow component is 315 s for the QD-1 and 225 s for the QD-2.

Utilizing a numerical approximation within the one-trap and one-recombination center model (OTOR), we have described the TSL curves quantitatively. The general-order kinetics model with discrete trapping levels unveils the dominance of $1^{st}$-order kinetic processes. A semi-analytical solution based on the Lambert $W$ function for the OTOR model was used, for the first time, to analyze thermally stimulated processes in QDs, which allows one to establish the retrapping ratio $R < 0.07$. The obtained values also confirm the low rate of recapture of charge carriers. The approach to the description of the TSL processes includes InP/ZnS samples' traps as having an discrete activation energy in the range of 25–31 meV. This stands for their common nature in the different-size nanocrystals tested. However, based on varying the holding time before the onset of linear heating, we gained experimental evidences, indicating possible distribution of parameters for active trapping centers in irradiated InP/ZnS QDs.

The results of this work demonstrate the prospects of using thermal activation spectroscopy approaches to study the properties of local defect states in the energy structure of colloidal quantum dots. Developing ideas about the origin of defects arising in QDs will help advance techniques for the targeted synthesis of nanocrystals with a high quantum yield of luminescence and adjustable temperature sensitivity. In the realm of sensor and optoelectronic technologies, all of the above said is an urgent task.